\documentclass[aps,prl,twocolumn,showpacs]{revtex4}
\usepackage{amssymb,amsmath}
\usepackage{bm, graphicx, amsmath}
\usepackage{bbm}
\usepackage[section]{placeins}

\usepackage[mathscr]{eucal}
\usepackage{graphicx}
\usepackage{color}
\usepackage{xcolor}

\usepackage{dutchcal}

\newcommand{\tr}{{\rm tr}\,}
  \newcommand{\omegap}{\omega'}

\newtheorem{lem}{Lemma}

\DeclareMathAlphabet{\mathpzc}{OT1}{pzc}{m}{it}

\def\blankpage{%
      \clearpage%
      \thispagestyle{empty}%
      \setcounter{equation}{0}
      \setcounter{figure}{0}
      \setcounter{page}{0}
      \null%
      \clearpage}
  
\begin{document}

  \title{
  Phase estimation with limited coherence 
  }
\author{D. Mu\~noz-Lahoz$^{1}$, J. Calsamiglia$^{1}$, J. A. Bergou$^{2,3}$, 
and E. Bagan$^{1}$}

\affiliation{$^{1}$F\'{i}sica Te\`{o}rica: Informaci\'{o} i Fen\`{o}mens Qu\`antics, Universitat Aut\`{o}noma de Barcelona, 08193 Bellaterra (Barcelona), Spain\\
$^{2}$Department of Physics and Astronomy, Hunter College of the City University of New York, 695 Park Avenue, New York, NY 10065, USA\\
$^{3}$Department of Mathematics, Graduate Center of the City University of New York, 365 Fifth Avenue, New York, New York 10016, USA
}

\begin{abstract} 

We investigate the ultimate precision limits for quantum phase estimation in terms of the coherence, ${\mathscr C}$, of the probe. For pure states, we give the minimum estimation variance attainable, ${\mathscr V}({\mathscr C})$, and the optimal state, in the asymptotic limit when the probe system size, $n$, is large. We prove that pure states are optimal only if ${\mathscr C}$ scales as $n$ with a sufficiently large proportionality factor, and that the rank of the optimal state increases with decreasing ${\mathscr C}$, eventually becoming full-rank. We show that the variance exhibits a Heisenberg-like scaling, ${\mathscr V}({\mathscr C})\sim a_{n}/{\mathscr C}^2 $, where $a_n$ decreases to $\pi^{2}/3$ as $n$ increases, leading to a dimension-independent relation.

\end{abstract}
\maketitle 

Resource theory made possible to quantify in a rigorous way what arguably is, along with entanglement, the most fundamental of all quantum-mechanical properties: coherence~\cite{baumgratz,levi,chitambar,winter,yadin,streltsov,ben dana}. Once coherence is recognized as a quantifiable resource, a natural question comes to mind: To what extent can central quantum-informational tasks be performed with a limited amount of coherence? Quantum state discrimination and estimation are two such tasks that arise in countless applications in quantum technologies. The relation between coherence and discrimination has been 
discussed in the literature. Hence, e.g., the role of coherence in wave-particle duality~\cite{wootters,greenberger,jaeger,englert,durr,englert2,coles1}, where phase and path discrimination are instrumental, has been firmly stablished~\cite{pati,bagan,baganGames,Sen}. Likewise, the measure of coherence named robustness of coherence (RoC) [see Eq.~(\ref{def of RoC}) below] has been shown to have a clear operational meaning in phase discrimination~\cite{pianiadessoprl,pianiadessopra}. 
As to estimation, there is a large body of work addressing the construction of (unspeakable~\cite{marvian2016}) coherence (or asymmetry~\cite{gour,marvian2014}) quantifiers from quantum metrology~\cite{zhang,streltsov,ben dana}.
However, to the best of our acknowledge, the precise link between RoC ---or other widespread quantifiers~\cite{streltsov}, such as the $l_1$ measure of coherence~\cite{baumgratz}--- and estimation problems~\cite{baganGill,Paris,Li} has not yet been tackled. Our aim is to take a step in this direction by addressing  phase estimation in quantum interferometry~\cite{holevo} (for a comprehensive review see also~\cite{kolodynski}) when the amount of coherence allowed to prepare the probe state is limited according to some prescribed quantifier. 
 
 To be more specific, we approach phase estimation from a Bayesian viewpoint, assuming no prior knowledge of the phase that we wish to estimate, which 
 possesses circular symmetry.
Thus, 
we use the Holevo's $2\pi$-periodicized variance~\cite{holevo,van Dam}, defined as \mbox{$
{\mathcal v}(\theta,\hat\theta)=4\sin^2[(\theta-\hat\theta)/2]$}, where~$\hat\theta$ stands for the guessed value of the true phase~$\theta$, 
to quantify the goodness of phase estimation [note that it respects circularity and approaches the standard square error/variance, $(\theta-\hat\theta)^2$, as $\hat\theta\to\theta$]. More precisely, we use the average 
\begin{equation}
{\mathscr V}(\rho)=\min \sum_{\chi}\int_0^{2\pi} {d\theta\over2\pi}
{\mathcal v}(\theta,\hat\theta_\chi)\tr[U(\theta)\rho U^\dagger(\theta)E_\chi],
\end{equation}
 where the minimization is over all measurements, i.e., positive operator valued measures $\{E_\chi\}$, and 
assignment rules, $\chi\mapsto\hat\theta_\chi$ (i.e., outcome post-processings). 
This formulation provides us with a global estimator that works for any possible value that the phase, imprinted to the probe state $\rho$ through the unitary transformation~$U(\theta)$, 
might have. 
Hereafter, we will refer to ${\mathscr V}(\rho)$ as variance for short.

It is well known~\cite{wiseman} that 
$
{\mathscr V}(\rho)=2-2\sum_{k=0}^{n-1} |\rho_{k\,k+1}|
$,
where~
$\rho$ is expressed in the eigenbasis, $\{|k\rangle\}_{k=0}^n$, of~$U(\theta)$, so that $U(\theta)=\sum_{k=0}^n{\rm e}^{-ik\theta}|k\rangle\langle k|$. The optimal probe state is pure, $\rho^*=|\phi^*\rangle\langle\phi^*|$, with
$|\phi^*\rangle=\sum_{k=0}^n \phi^*_k|k\rangle$ and
\begin{equation}
\phi^*_k=\sqrt{2\over n+2}\sin\left({k+1\over n+2}\pi\right)> 0.
\label{c*}
\end{equation}
Let ${\mathscr V}^*$ and ${\mathscr C}^*$ stand respectively for the absolute (unconstrained) minimum of $\mathscr V$ and the minimum 
RoC that is required to attain~${\mathscr V}^*$. It is known that~\cite{wiseman}
\begin{equation}
{\mathscr V}^*=4\sin^2\left({\pi\over 2n+4}\right)\sim{\pi^2\over n^2},\label{H*}
\end{equation}
where $\sim$ stands for asymptotic behavior as $n\to\infty$ 
(hereafter asymptotics will always refer to this limit).
We recall that RoC is defined as~\cite{pianiadessoprl}
\begin{equation}
{\mathscr C}_R(\rho):=\min_{\tau}\left\{s \ge0: {\rho+ s\tau\over s+1}\in {\mathscr I}\right\},
\label{def of RoC}
\end{equation}
where $\tau$ are physical states and  ${\mathscr I}$ is the set of incoherent states, i.e., density matrices diagonal in the reference basis under consideration. In our case, this is 
the eigenbasis of $U(\theta)$. 
It is known that for pure states RoC and the $l_1$-norm of coherence, ${\mathscr C}_{l_1}(\rho)=\sum_{k\not=k'}^n|\rho_{k\,k'}|$, coincide~\cite{pianiadessoprl}, thus, one has~\mbox{${\mathscr C}_R(\phi)={\mathscr C}_{l_1}(\phi)=(\sum_{k=0}^n \phi_k)^2-1$} for any generic pure state $|\phi\rangle=\sum_{k=0}^n\phi_k|k\rangle$. In particular, for $|\phi^*\rangle$, a straightforward calculation leads to 
\begin{equation}
{\mathscr C}^*={2\over n+2}\cot^2\!\left({\pi\over 2n+4}\right)-1\sim {8n\over\pi^2}.
\end{equation}
We point out that ${\mathscr C}^*< {\mathscr C}^{\rm max}_n:=n$, where  ${\mathscr C}^{\rm max}_n$ is the maximum coherence one can possibly have in a Hilbert space of dimension~$n+1$. In other words, maximally coherent probe states are not required for optimal phase estimation in the Bayesian framework. Note also that ${\mathscr V}^*$ scales as $1/n^2$ (Heisenberg-limited scaling), and ${\mathscr C}^*$ as $n$.

We now ask ourself what is the minimum value of the variance that can be achieved if the probe has bounded coherence, \mbox{${\mathscr C}_R(\rho)\le{\mathscr C}$}, where ${\mathscr C}$ is a given constant. In this case, what is the optimal probe state? These are meaningful questions since coherence is a valuable resource and often limited in physical implementations. These questions are also meaningful from a fundamental perspective and they are
answered below. 

Our results can be summarized as follows. For pure states, and in the asymptotic limit, we give analytic expressions of the minimum variance as a function of the available amount of coherence ${\mathscr C}$. We prove that pure states are optimal if  the coherence ratio $c:={\mathscr C}/{\mathscr C}^*$ is above~$82\%$. Below this ratio, pure states are suboptimal, which may come as a surprise. 
%
We also show that for \mbox{$62\% < c <82\%$} the optimal state is actually a rank~2 mixed state, and we 
give its expression in the asymptotic limit.
In the low coherence regime ($c<62\%$), we introduce a simple model of the optimal state that 
provides a very tight upper bound (hence a good approximation) to the minimum variance and proves the suboptimality of pure states. Finally, with independence of the probe-state size $n$, we find the ultimate phase estimation precision achievable with limited coherence. The minimum variance is shown to scale as~\mbox{$1/{\mathscr C}^2$}. 


{\em Pure probe states.} We first focus in probes of (asymptotically) large dimensionality, prepared in a pure state. We write the variance as ${\mathscr V}=2S/(n+2)^2$, where~\cite{action}
\begin{equation}
S=\sum_{k=-1}^n \epsilon\, {1\over2} \left({\phi_{k+1}/\sqrt\epsilon-\phi_k/\sqrt\epsilon \over \epsilon }\right)^2,
\label{free action}
\end{equation}
and we have defined \mbox{$\epsilon=1/(n+2)$}, \mbox{$\phi_{-1}:=\phi_{n+1}:=0$}. Asymptotically, $\epsilon k$ effectively becomes a continuous variable, $t\in[0,1]$, the term in the parenthesis becomes a derivative with respect to $t$, and the sum 
approaches
\mbox{$S\!=\!\int_0^1 (1/2)\dot x^2(t)dt$}, which we identify as the ``action of a free particle'' whose position is given by $x(t)$, with $x(\epsilon k)=\phi_{k-1}/\sqrt\epsilon$, $k=1,2,\dots,n+1$, and the boundary conditions~$x(0)=x(1)=0$. 
We need to introduce Lagrange multipliers to take into account the normalization of the probe state, $\sum_{k=0}^n \phi_k^2=1$; the maximum amount of coherence allowed, 
$\sum_{k=0}^n \phi_k\le({\mathscr C}+1)^{1/2}$; 
and the non-negativity of the coefficients, $\phi_k\ge0$. Adding these constraints, the ``free action", given above, becomes
\begin{multline}
S =  \int_0^1   \Big\{ {1\over 2}\dot x^2(t) - {\omega^2\over2} \left[x^2(t) - 1\right]\\
 - \mu^2\left[x(t)-l \right]+ \sigma(t) x(t) \Big\},
 \label{S + lagrange}
\end{multline}
in the asymptotic limit.
Here the parameter $l$ is related to the maximum amount of coherence allowed,~${\mathscr C}$, through \mbox{$l=[({\mathscr C}+1)/(n+2)]^{1/2}$}, the multipliers  $\omega^2$, $\mu^2$ and~$\sigma(t)$ are non-negative and  
$\mu\!\int_0^1[x(t)\!-\!l]dx=0$, \mbox{$\sigma(t)x(t)=0$} (slackness conditions). 

\begin{figure}[ht]
\includegraphics[scale=.145]{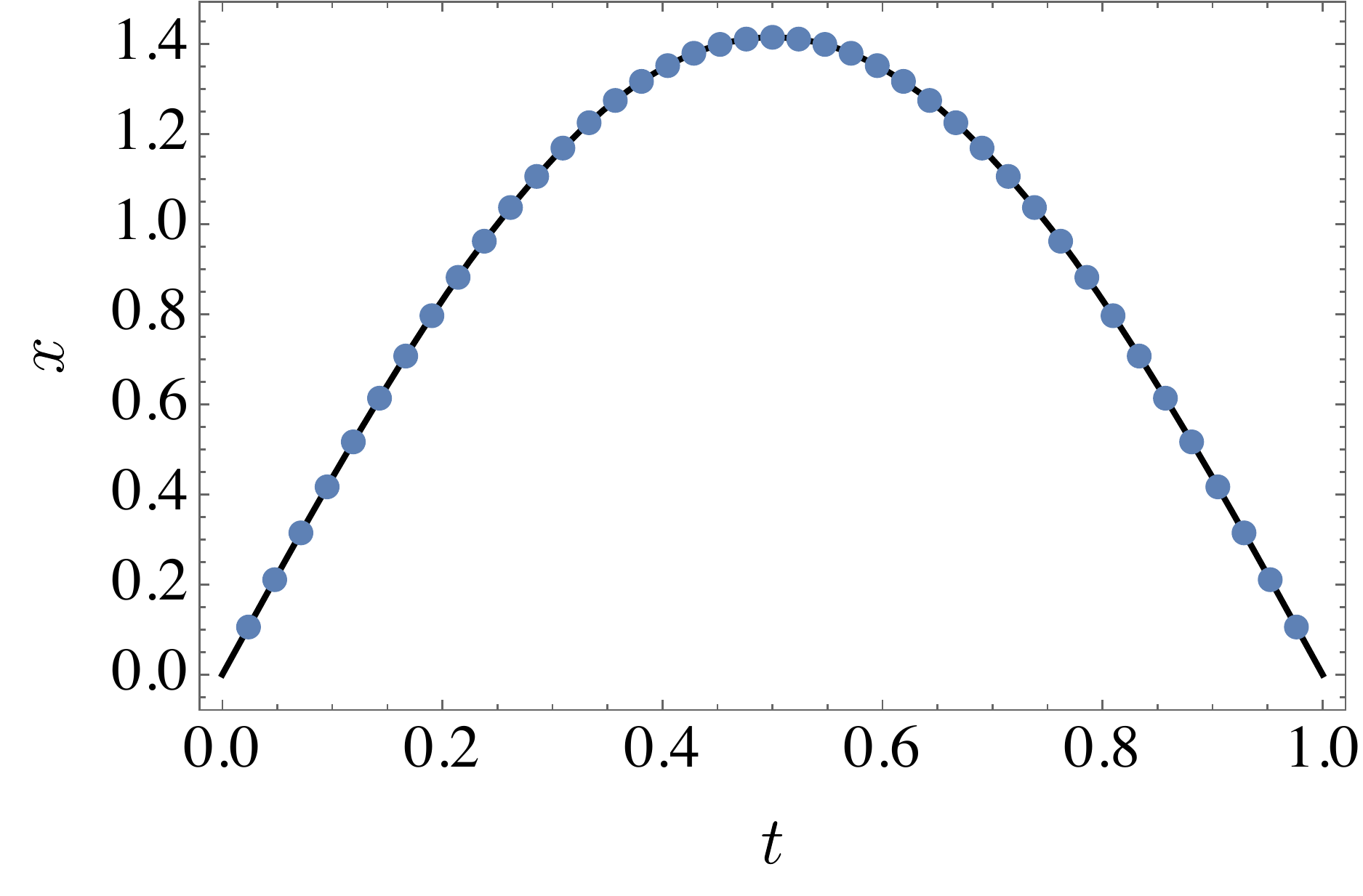}
\includegraphics[scale=.145]{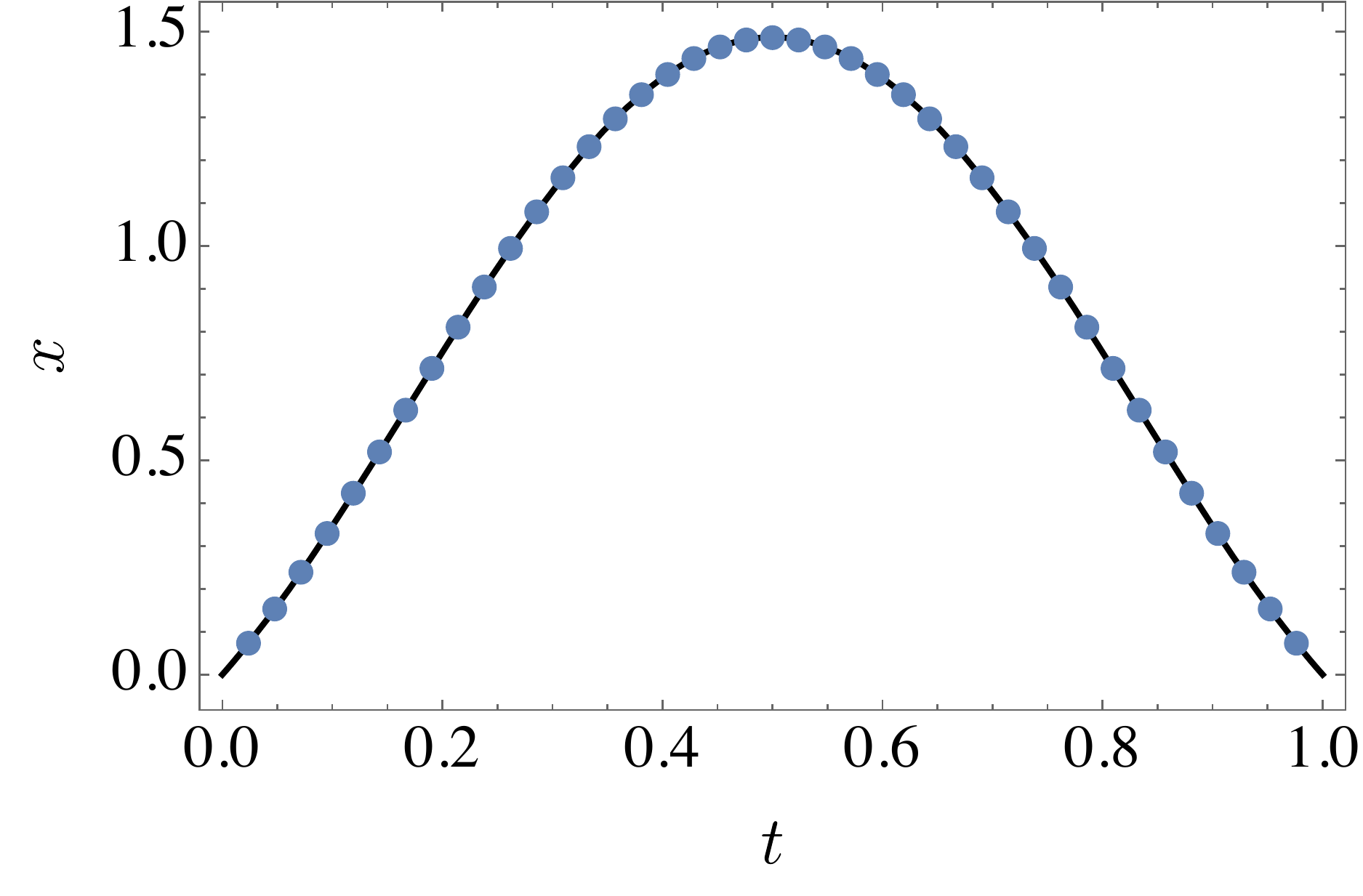}
\includegraphics[scale=.145]{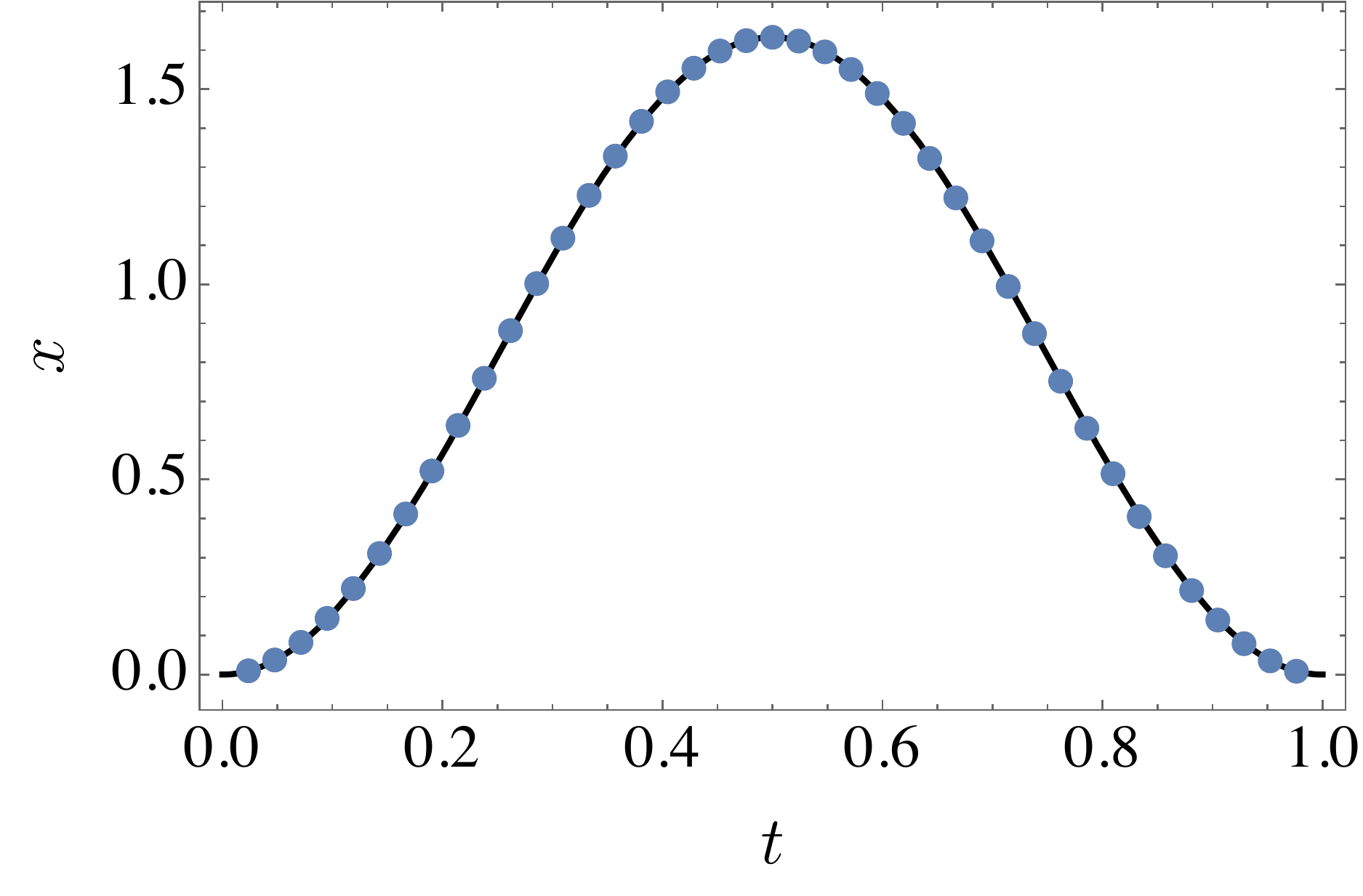}
\includegraphics[scale=.145]{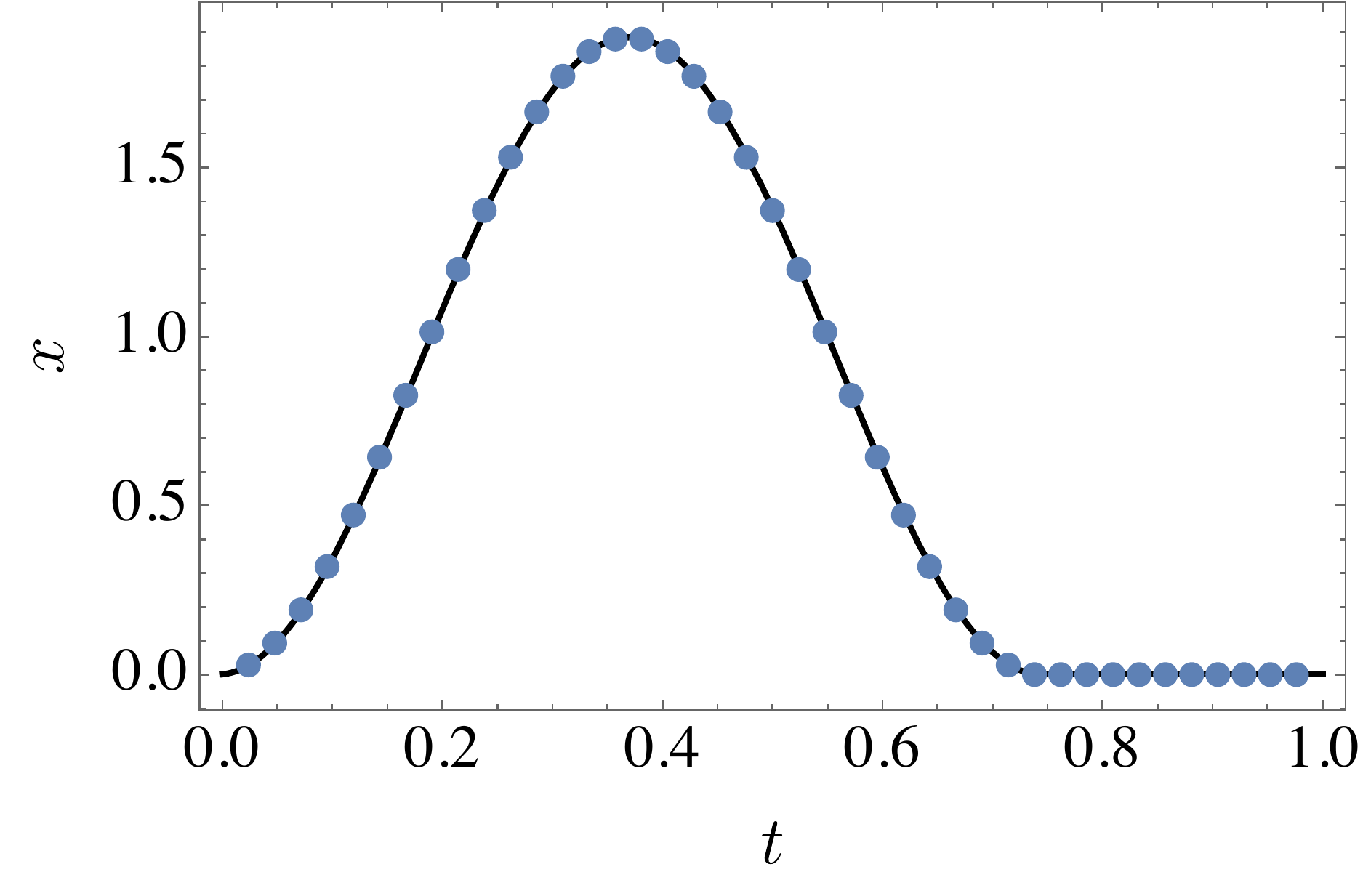}
\includegraphics[scale=.145]{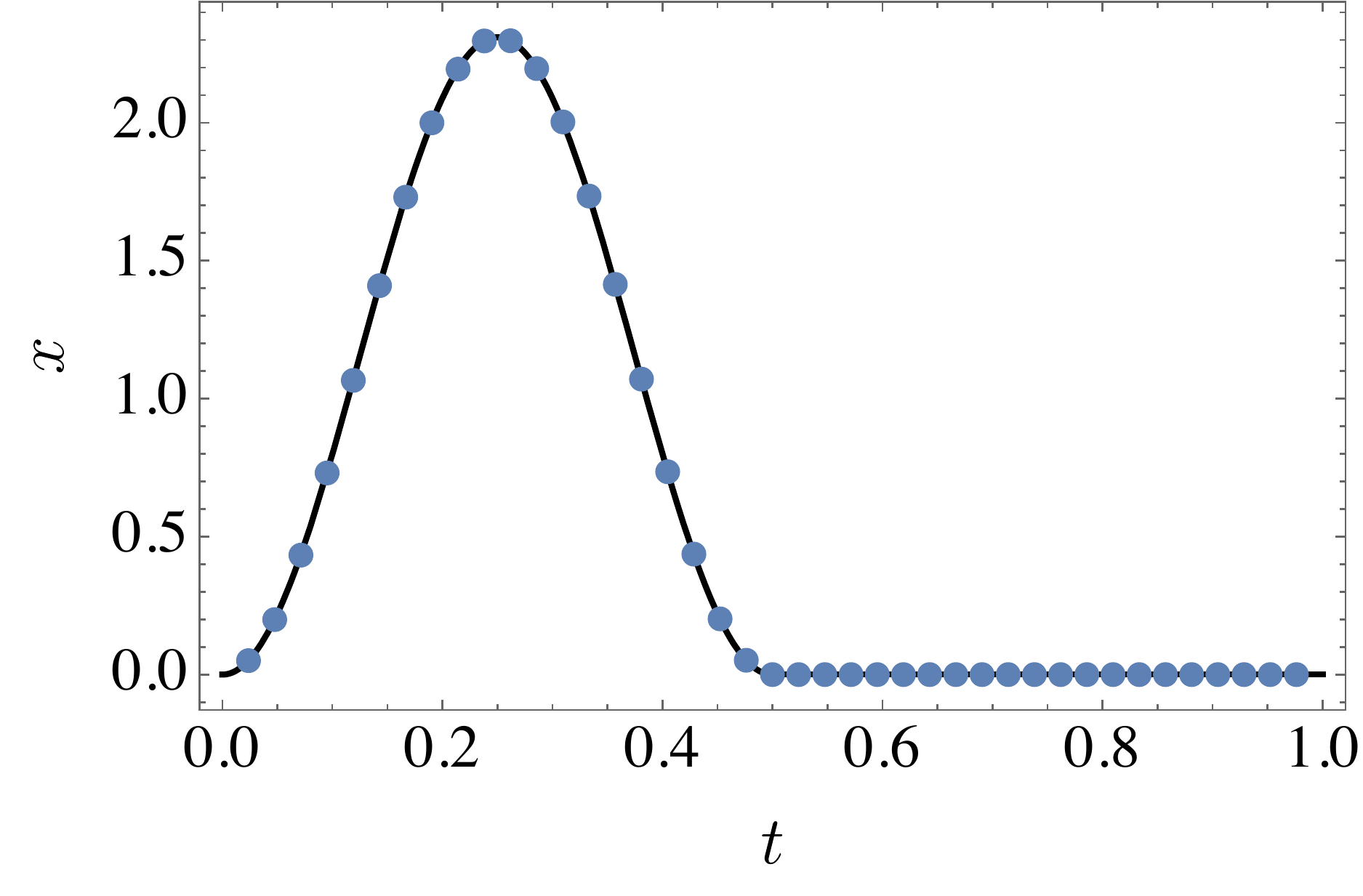}
\includegraphics[scale=.145]{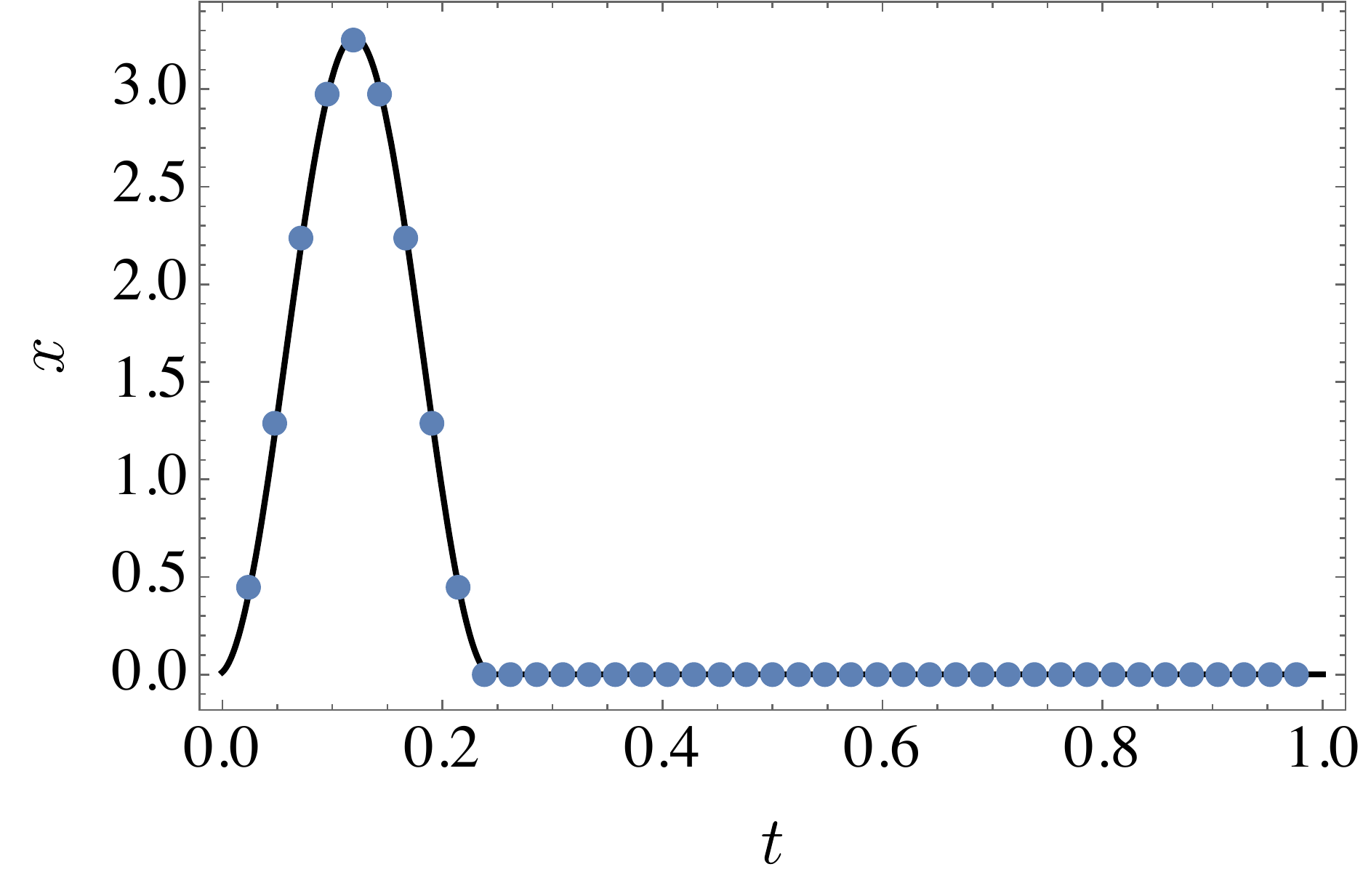}
\caption{The solid lines show the shape 
of the optimal asymptotic pure state, Eqs.~(\ref{x omega}) and~(\ref{x a}), for various values of $\omega$ (top row) and $a$ (bottom row). The dots are the points $(\epsilon k,\phi_{k-1}/\sqrt\epsilon)$, 
$k=1,2,\dots,n+1$, for $n=40$, where the coefficients~$\phi_k$ are computed by numerical optimization. In the top row $\omega=\pi,\,3\pi/2,\,2\pi$ ($c=1.00,\,0.95,\,0.82$, respectively). In the bottom row $a=3/4,\,1/2,\,1/4$ ($c=0.62,\,0.41,\,0.21$, respectively).
} 
\label{fig:3}
\end{figure}

The Euler-Lagrange equation gives the ODE that $x(t)$ must fulfill to minimize $S$, and thus ${\mathscr V}$ itself: 
\begin{equation}
\ddot x+\omega^2 x+\mu^2=0,\;\mbox{if $x>0$ ($\sigma=0$)}.
\label{EDO}
\end{equation}
For sufficiently large $l$ (namely, $ {\mathscr C}\ge{\mathscr C}^*$) the constraint $\int_0^1 x(t)dt\le l$ is not active and thus $\mu=0$. Eq.~(\ref{EDO}) becomes the ODE of a harmonic oscillator and the solution that satisfies the remaining constrains and the boundary conditions 
is $x(t)=\sqrt2\sin(\pi t)$, which is the asymptotic (continuous) version of $\phi^*_k$ given in Eq.~(\ref{c*}). 
When the constraint is active ($\mu^2>0$), the solution of Eq.~(\ref{EDO}) that satisfies the normalization and limited-coherence constraint, along with the boundary conditions, is
\begin{equation}
x(t)=2A(\omega)\sin\left({\omega t\over2}\right)\sin\left[{\omega(1-t)\over2}\right],
\label{x omega}
\end{equation}
where the normalization factor, $A(\omega)$, can be written as
$
A(\omega)=[1+(\cos\omega)/2-(3\sin\omega)/(2\omega)]^{-1/2}
$ and $\omega$ is not arbitrary, but satisfies 
\begin{equation}
l=f(\omega):=A(\omega)\left[{2\over\omega}\sin\left({\omega\over2}\right)-\cos\left({\omega\over2}\right)\right].
\label{f(w)}
\end{equation}
All the required conditions are satisfied if \mbox{$\pi\le\omega\le2\pi$}, and $\lim_{\omega\to\pi\downarrow} x(t)=\sqrt2\sin(\pi t)$, as it should be.
The solution in Eq.~(\ref{x omega}) is shown in the top row of Fig.~\ref{fig:3} for various values of $\omega$ (solid line) along with the (scaled) coefficients~$\phi_k$, obtained by numerical optimization (points). Note the perfect agreement between analytical and numerical results. 
%
%

 Substituting now Eq.~(\ref{x omega}) in Eq.~(\ref{S + lagrange}), we obtain the minimum ``action": \mbox{$S=S(\omega):=(\omega^2/4)A^2(\omega)[1\!-\!(\sin\omega)/\omega]$}.  At the lower end of the $\omega$ interval, $\omega=\pi$, one can check that the optimal unconstrained variance ${\mathscr V}^*$ in Eq.~(\ref{H*}) is recovered. As a result of this derivation, we have managed to express ${\mathscr V}$ and ${\mathscr C}$ [through $S$ and $l$] in terms of the parameter~$\omega\in[\pi,2\pi]$.
 
 However, if~$\omega>2\pi$, the solution in Eq.~(\ref{x omega}) does not satisfy $x(t)\ge0$ in the entire interval $[0,1]$. We thus must find an alternative solution. There must exist an interval $(a,1]\subset[0,1]$ (the so-called coincidence set) where 
 the alternative solution satisfies $x(t)=0$ [and~$\sigma(t)>0$]. In the complementary region, $[0,a]$ the solution that satisfies the requirements (including now differentiability at~$t=a$) can be shown to be
\begin{equation}
x(t)=\sqrt{8\over 3a}\sin^2\left({\pi t\over a}\right),\;  0\le t\le a\label{x a}
\end{equation}
(and, as mentioned, zero elsewhere)~\footnote{It is apparent from the asymptotic form of $S$ in Eq.~(\ref{S + lagrange}) 
that the solutions are degenerate in the regime governed by the parameter~$a$: the shift $x(t)\mapsto x(t-t_0)$, where $t_0\in[0,1-a]$, produces a new solution with the same value of $S$. This degeneracy takes also place for finite $n$ provided the coherence is low enough.}. In this regime one has $l=(2a/3)^{1/2}$ and $S=2\pi^2/(3a^2)$. From the expression of $l$, we see that the support of the optimal pure state begins to shrink as soon as 
the coherence ratio $c$ falls below \mbox{$\pi^2/12\approx 82.2\%$} (see Fig.~\ref{fig:3}), so, effectively, 
the optimal
pure state has smaller dimensionality, \mbox{$n'\sim (12c/\pi^2) n<n$}, which explains why pure states, as we will find below, are not optimal in this regime. 
%
%
The bottom row of Fig.~\ref{fig:3} contains three plots of $x(t)$ in Eq.~(\ref{x a}), showing perfect agreement with numerical data. 

\begin{figure}[ht]
\includegraphics[scale=.4]{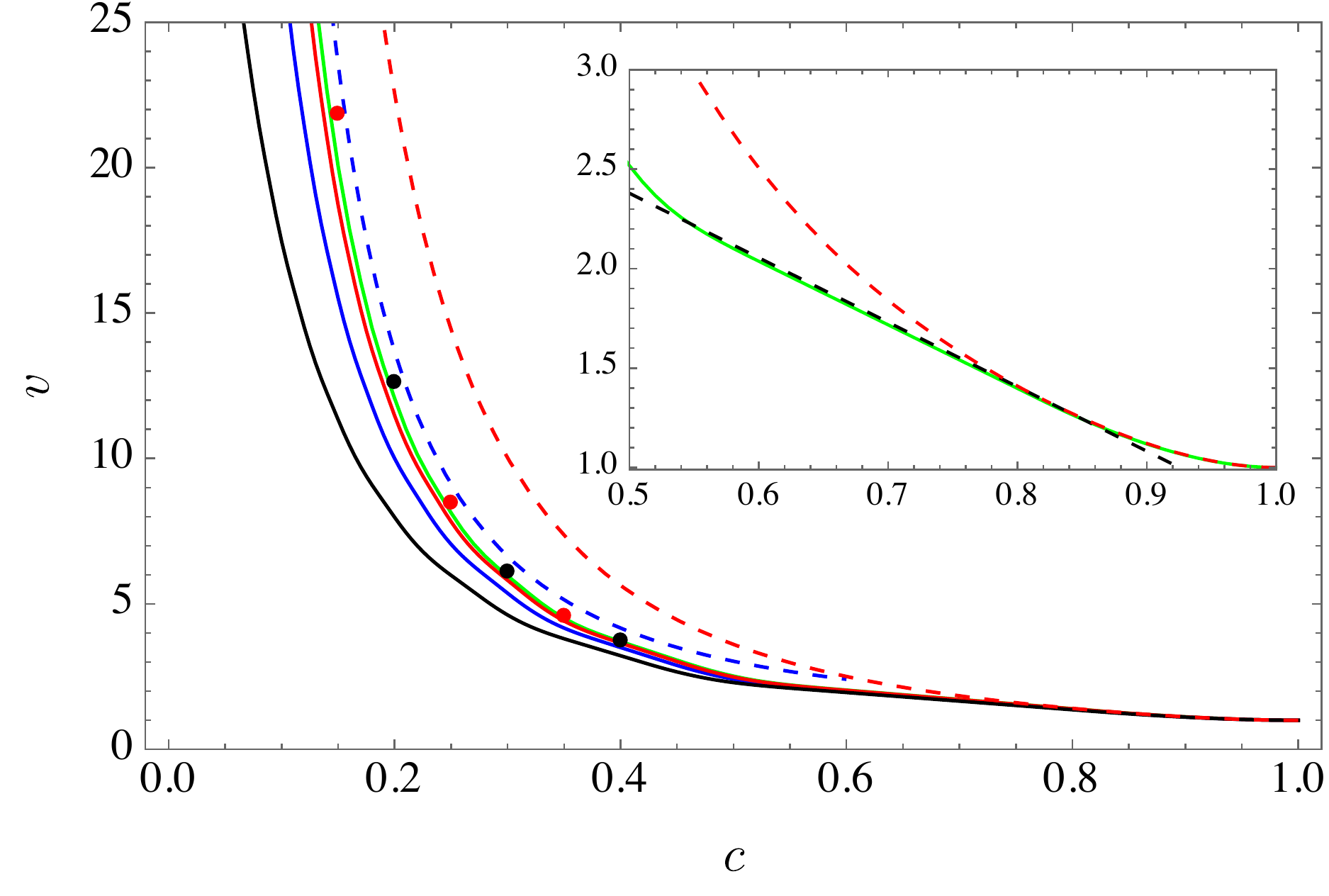}
\caption{$v$ vs $c$ for mixed probe states. The black, blue, red and green solid lines correspond to $n=15,\,30,\,60$, and $90$, respectively. 
The black (red) dots correspond to $n=160$ ($n=220$). Both, lines and dots have been obtained by numerical optimization.
The red dashed line is the asymptotic pure-state curve. 
The blue dashed line is the bound in Eq.~(\ref{hol bound}) and it is seen to prove that pure states are not optimal if~\mbox{$c\lessapprox 60\%$}. The black dashed line (inset) shows the lower bound $4(1-{8c/\pi^2})\le v$, attained by the mixed state in~Eq.~(\ref{the rho guess 2}) in the range $\pi^2/16\!\le\! c\!< \!\pi^2/12$.
} 
\label{fig:4}
\end{figure}

Fig. \ref{fig:4} (red dashed line in both main figure and inset) shows a plot of ${\mathscr V}$ vs $ {\mathscr C}$, obtained from the expressions of~$S$ and~$l$ that we have derived.  We choose to normalize these magnitudes to their corresponding optimal values, ${\mathscr V}^*$ and ${\mathscr C}^*$, i.e., we use $c:= {\mathscr C}/{\mathscr C}^*$ (introduced above) and $v:={\mathscr V}/{\mathscr V}^*$ to remove the~$n$ dependence from the plot. 
%
%


{\em General probe states.} Optimal phase estimation with limited RoC can be easily cast as a semidefinite program~(SDP)~\cite{boyd}:
\begin{equation}
{\mathscr V}=\min_{\rho,\tau} \tr(H\rho)\;\mbox{$\begin{array}{c}{\rm subj.}\\ {\rm to}\end{array}$}\; \left\{\begin{array}{l}\rho+{\mathscr C} \tau\in{\mathscr I}\; \mbox{(diagonal),}\\ \rho\ge0,\;\tau\ge0,\\ \tr\rho=\tr\tau=1,\end{array}\right.
\label{primal SDP}
\end{equation}
where  we have introduced the $(n\!+\!1)\times(n\!+\!1)$ matrix \mbox{$H=\operatorname{tridiag}(-1,2,-1)$} (the discrete 1-D Laplace operator), to write the variance as \mbox{${\mathscr V}(\rho)=\tr(H\rho)$}, 
and the first constraint enforces the condition ${\mathscr C}_R(\rho)\le{\mathscr C}$ [see Eq.~(\ref{def of RoC})].

SDPs can be very efficiently solved numerically. Fig.~\ref{fig:4}  shows plots of $v$ vs $c$ for various values of $n$ (solid lines and dots). The (rounded) kinks in the curves call for a comment. 
A~close look at the structure of $\rho$ provided by the SDP solutions 
shows that its rank  increases by one at each subsequent kink as $c$ decreases, eventually becoming a full rank incoherent state~\footnote{We note that the curves move up with increasing values of~$n$, which may look contradictory. Recall, however, that we are plotting a ratio of variances, rather than the variance itself. The plots show that in percentage terms, and over the base value~${\mathscr V}^*$ (which decreases as $1/n^2$), the percentage value increases with $n$. 
}. 
The first occurrence takes place at $c=\pi^2/12$, where the (pure state) solution $x(t)$ changes regime, from Eq.~(\ref{x omega}) to Eq.~(\ref{x a}), and pure states cease to be optimal.

{\em Optimality/suboptimality of pure states.} Fig.~\ref{fig:4} clearly shows that in our constrained optimization problem, mixed states perform better than pure states when the ratio $c$ is less than about 82\%.  We prove in the supplemental material~\cite{supp} that this advantage persists in the asymptotic limit. Beyond the numerical evidence provided by this plot, we also prove in full detail that pure states are indeed optimal for larger values of $c$. 
The proof consists in finding a feasible solution of the dual SDP ---and hence a lower bound to the optimal solution--- that is attained by our pure state result in Eq.~(\ref{x omega}). 
%
This can be achieved provided~$c$ is in the range $1\ge c\ge\pi^2/12\approx 82.2\%$.

Outside this range, this very same feasible dual solution 
leads to the asymptotic lower bound
$
4(1-{8c/\pi^2})\le v
$
 (dashed black straight line in the inset of Fig.~\ref{fig:4}),
that 
is {\em not} attainable with pure states (dashed red curve), except at the single point $c=\pi^2/12$.
In~\cite{supp}, we show that this lower bound is attained, as $n\to\infty$, by the rank two probe state
\begin{equation}
\rho={3l^2\over2}|\phi_{\rm even}\rangle\langle\phi_{\rm even}|\!+\!\left(\!1\!-\!{3l^2\over2}\right)|\phi_{\rm odd}\rangle\langle\phi_{\rm odd}|,
\label{the rho guess 2}
\end{equation}
for $\pi^2/16\!\le\! c\!< \!\pi^2/12$ ($\sqrt{1/2}\!\le\! l\!\le\! \sqrt{2/3}$, respectively), where 
\begin{align}
|\phi_{\rm even}\rangle&=\sqrt{8\over3(n+2)}\;\sum_{k=0}^n \sin^2\left(\pi{k+1\over n+2}\right)|k\rangle,
\label{even state}\\
|\phi_{\rm odd}\rangle&=\sqrt{2\over n+2}\;\sum_{k=0}^n \sin\left(2\pi{k+1\over n+2}\right)|k\rangle.
\label{odd state}
\end{align}
Here, even/odd refers to the obvious symmetry of these states.
Hence, the state in Eq.~(\ref{the rho guess 2}) is optimal (or in other words, pure states are suboptimal) in this range of~$c$. As a by-product, we have the exact analytical solution of our optimization problem for $c\ge 61.7\%$ [Eqs.~(\ref{x omega}) and~(\ref{the rho guess 2})].

{\em Low coherence regime.} Because of the increasing rank of $\rho$, it becomes much harder to 
carry out the detailed analysis above for lower values of the coherence. Instead, we introduce a simple ansatz for the probe state $\rho$ that is shown to give a very tight upper bound to the asymptotic optimal curve~$(c,v)$ and 
proves the suboptimality of pure states when \mbox{$c\lessapprox 61\%$}. It is given by
\begin{multline}
\tilde\rho(s)={2\over n+2}\sum_{k,k'=0}^n\sin\left({k+1\over n+2}\pi\right)\sin\left({k'+1\over n+2}\pi\right)\times\\
\operatorname{sinc}\left({k-k'\over s+1}\pi\right)|k\rangle\langle k'|,\quad s\ge 0,
\end{multline}
where we use the common definition \mbox{$\operatorname{sinc} x:=(1/x)\sin x$}.
The operator $\tilde\rho(s)$ is a proper physical state, as it is the result of (partially) twirling the optimal state in Eq.~(\ref{c*}): $\tilde\rho(s)={\mathcal T}_s(\rho^*)$, where ${\mathcal T}_s$ is the (trace-preserving) channel \mbox{${\mathcal T}_s(\sigma):=\int_{-\alpha}^\alpha U(\phi)\sigma U^\dagger(\phi)\, d\phi/(2\alpha)$}, with $\alpha=\pi/(s+1)$. As $s$ approaches zero, ${\mathcal T}_s$ becomes fully decohering, and ${\mathscr C}_R(\tilde\rho(0))=0$. If, on the other hand,~$s$ is of order $n$, one expects that coherence will not decrease significantly.  
%
%
%
It turns out that~$s$ is a good approximation to ${\mathscr C}_R(\tilde\rho(s))$ if $s<n/2$, 
but the precise functional relation between these two quantities seems hard to establish and we will content ourselves with finding a bound.

To this end, let us consider the trace-one operator
\begin{equation}
\tau={2(s+1)\over s(n+2)}\sum_{k=0}^n\sin^2\left({k+1\over n+2}\pi\right)|k\rangle\langle k|-{\tilde\rho(s)\over s}.
\end{equation}
We note that $\tau=(s+1)(\openone-M)\circ\rho^*/s$, where~$\circ$ stands for the Hadamard product in our reference basis, $\{|k\rangle\}_{k=0}^n$, and
\begin{equation}
M=\sum_{k,k'=0}^n{ \operatorname{sinc}\left({k-k'\over s+1}\pi\right)\over s+1}|k\rangle\langle k'| .
\end{equation}
The matrix $M$ is a discrete version of the Dyson (or sinc) kernel, widely used in signal processing, random matrices and numerical analysis~\cite{slepian,delsarte}. It is known that the maximum eigenvalue of $M$ is one, from which it follows that $\tau\ge0$  (we recall that the Hadamard  product of positive semidefinite matrices is a positive semidefinite matrix), and thus $\tilde\rho(s)+s\tau\in{\mathscr I}$
%
(i.e., it is incoherent). We conclude from Eq.~(\ref{def of RoC}) that
the RoC of $\tilde\rho(s)$ is bounded~as~${\mathscr C}\le s$.
%
%
By a straightforward calculation,  the variance of our ansatz satisfies
%
\begin{equation}
{\mathscr V}(\tilde\rho(s))\le 2-(2-{\mathscr V}^*)\operatorname{sinc}\left({\pi\over {\mathscr C}+1}\right).
\label{H john 0}
\end{equation}
%
%
Taking the limit $n\to\infty$ we obtain the asymptotic curve
\begin{equation}
v=1+{\pi^4\over192 c^2},
\label{hol bound}
\end{equation}
%
represented by the blue dashed line in Fig.~\ref{fig:4}. By construction, it is an upper bound to the optimal asymptotic curve~$(c,v)$, and rigorously proves that for low coherence ($c\lessapprox 62\%$) pure probe states are suboptimal. Numerical results, also shown in Fig.~\ref{fig:4}, reveal that actually Eq.~(\ref{hol bound}) gives a very good approximation to the optimal asymptotic curve if $c\le50\%$. The approximation is seen to improve with decreasing $c$.

%

\begin{figure}[thb]
\includegraphics[scale=.4]{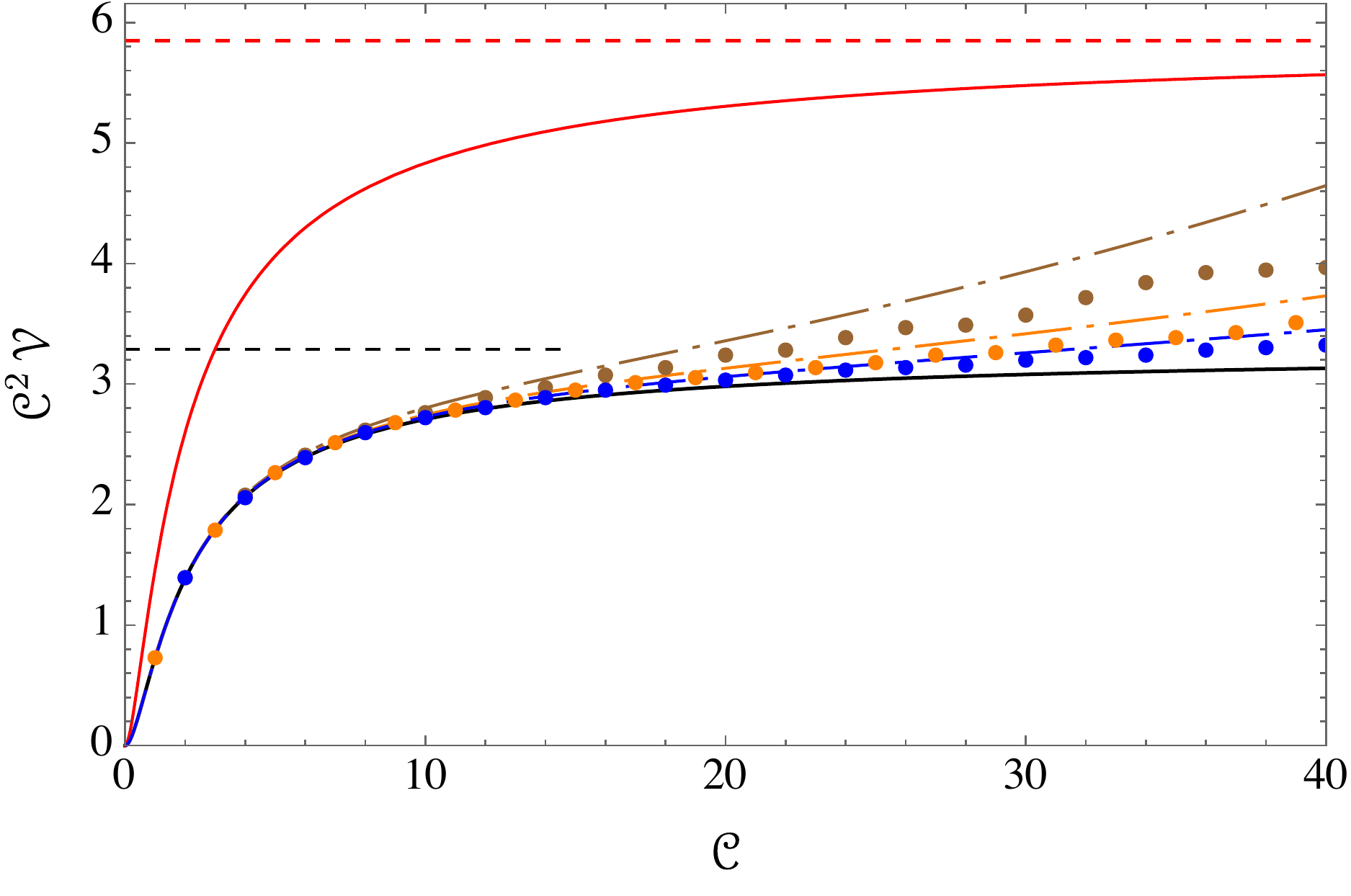}
\caption{
The graphs of ${\mathscr C}^2{\mathscr V}_{\rm lim}({\mathscr C})$ (black solid line) and ${\mathscr C}^2{\mathscr V}_{\rm lim}^{\rm pure}({\mathscr C})$ (red solid line). Here, the product ${\mathscr C}^2{\mathscr V}$ is used instead of~${\mathscr V}$ to better exhibit the convergence of the numerical data (colored points) to the analytic curves and the large~${\mathscr C}$ behavior (dashed lines) of the variance (i.e., its scaling as $1/{\mathscr C}^2$). The analogous graphs derived from the $n$-dependent bound in Eq.~(\ref{H john 0}) are also plotted (dot-dashed lines). Brown, orange, and blue correspond to $n=100,\ 160$ and $220$, respectively.
} 
\label{fig:60}
\end{figure}

{\em Dimension independent bounds.} So far we have assumed that the value of $n$, the size of the probe system, is given and the optimization has been carried out for such fixed values. Our aim is now to lift this condition, so that coherence is the only restriction placed on the estimation problem. From 
Eq.~(\ref{H john 0}) and the short discussion that follows Eq.~(\ref{hol bound}) one immediately has
\begin{equation}
{\mathscr V}\lessapprox {\mathscr V}_{\rm lim}({\mathscr C}):=2-2\operatorname{sinc}\left({\pi\over {\mathscr C}+1}\right),
\label{H john 2}
\end{equation} 
which is obtained for $n\to\infty$ and $\rho$ of full rank. The product $ {\mathscr C}^2{\mathscr V}_{\rm lim}({\mathscr C})$  is plotted in Fig.~\ref{fig:60} (black solid lines) along with some numerical data (points) and the finite-$n$ upper bound in Eq.~(\ref{H john 0}) (dot-dashed lines), where brown, orange and blue correspond to $n=100$, $160$, and $220$, respectively. Fig.~\ref{fig:60}  provides strong evidence that 
${\mathscr V}_{\rm lim}({\mathscr C})$ is also a lower bound and, thus, the exact asymptotic expression of ${\mathscr V}({\mathscr C})$. 
The analog of Eq.~(\ref{H john 2})  for pure states can be easily derived from the expressions of $S$ and~$l$ after Eq.~(\ref{x a}):
${\mathscr V}^{\rm pure}_{\rm lim}({\mathscr C})= (16/27)\pi^2/({\mathscr C}+1)^2$ (red solid line in Fig.~\ref{fig:60}). Once again, the plot illustrates that pure states are not optimal.
For large values of ${\mathscr C}$,  $ {\mathscr V}_{\rm lim}({\mathscr C})$ has the asymptotic form
\begin{equation}
{\mathscr V}_{\rm lim}({\mathscr C})\sim {\pi^2/3\over {\mathscr C}^2}
\end{equation} 
(black dashed line; in red for pure states). This scaling of the  (ultimate)  variance resembles that of Heisenberg-limited quantum interferometry/metrology, with coherence playing the role of the probe-state dimension.

\begin{acknowledgments}
\emph{Acknowledgments}. We are grateful to A.~Winter, M.~Garc\'{\i}a D\'{\i}az and M. Hillery for discussions. This research was supported by  the
Spanish Agencia Estatal de Investigaci\'on,~project PID2019-107609GB-I00, and the~Ge\-ne\-ra\-li\-tat~de~Ca\-ta\-lu\-nya CIRIT, contract  2017-SGR-1127. JC also acknowledges support from ICREA Academia award.
\end{acknowledgments}

\newpage
\blankpage

\appendix*

\renewcommand{\theequation}{\arabic{equation}}

\section{SUPPLEMENTAL MATERIAL}

\subsection{Dual SDP}

In this section, we derive the expression of the semidefinite program (SDP) dual to that in Eq.~(\ref{primal SDP}) in the main text (MT).
From this equation 
we can write down the~``Lagrangian":
\begin{multline}
L=\tr(H\rho)+r\left(\tr\rho-1\right)+y\left(\tr\tau-1\right)\\
+\tr\left[\left(\rho+{\mathscr C}\tau\right)Z\right]-\tr(\rho R)-\tr(\tau T),
\label{supp primal}
\end{multline}
where $Z$ is a hermitian matrix 
with zeros in its main diagonal. It consists  of the Lagrange multipliers that enforce the constrain \mbox{$\rho+{\mathscr C}\tau\in{\mathscr I}$}. 
Likewise, $y,\, r\in{\mathbb R}$ are also Lagrange multipliers, and the matrices~$R$ and~$T$ are positive semidefinite. Eq.~(\ref{supp primal}) can be rearranged~as
\begin{multline}
L=\tr\left[\rho\left(H+r\openone+Z-R\right)\right]\\
+\tr\left[\tau\left(y\openone+{\mathscr C}\,Z-T\right)\right]-r-y .
\label{L inter}
\end{multline}
With the choice
\begin{align}
R&=H+r\openone+Z,\\
T&=y\openone+{\mathscr C}\,Z,
\end{align}
one gets rid of the matrices $\rho$ and $\tau$ in Eq.~(\ref{L inter}) and the Lagrangian becomes \mbox{$L=-y-r$}. Hence, the dual program can be cast as
\begin{equation}
{\mathscr V}=-\min_{y,r,Z}(y+r)\quad\mbox{$\begin{array}{c}{\rm subj.}\\ {\rm to}\end{array}$}\;
\left\{
\begin{array}{rcl}
y\openone+{\mathscr C}\,Z\!\!&\ge&\!\!0,\\[.25em]
H+r\openone+Z\!\!&\ge&\!\!0,\\[.25em]
\Delta(Z)\!\!&=&\!\!0,
\end{array}
\right.
\label{supp DSDP}
\end{equation} 
where $\Delta$ is the dephasing operation, 
\begin{equation}
\Delta(\rho):=\sum_{k=0}^n|k\rangle\langle k|\rho|k\rangle\langle k|.
\end{equation}
In is more common (but, obviously, equivalent) to introduce the variables $y$ and $r$ with opposite sign and express the minimum variance as ${\mathscr V}=\max_{y,r,Z}(y+r)$.  It is then (even more) apparent that any feasible solution of the primal (dual) problem is an upper (lower) bound to ${\mathscr V}$. 

\subsection{Lower bound and optimality of pure probe states}

Here we prove the optimality of pure states. By choosing a particular simple form for $Z$, we will find a feasible solution of the dual SDP, Eq.~(\ref{supp DSDP}). This will provide a lower bound to $\mathscr V$ . We will then check its attainability with pure states. 
Attainability will hold if the allowed coherence is greater than $\pi^2/12\approx 82.2\%$ of ${\mathscr C}^*$. 

Let us assume that $Z=z[|\Psi^+\rangle\langle\Psi^+|-\openone]$, where $z\ge0$ and $|\Psi^+\rangle=\sum_{k=0}^n|k\rangle$ is the (unnormalized) maximally coherent pure state. The eigenstates of $y\openone+{\mathscr C}Z$ are: (i)~the state~$|\Psi^+\rangle$ itself, and (ii)~ any orthonormal basis of the $n$-dimensional subspace orthogonal to $|\Psi^+\rangle$. The corresponding eigenvalues are (i)~$y+n z {\mathscr C}$ and (ii)~\mbox{$y-z {\mathscr C}$}, where the latter has multiplicity~$n$. Thus, we have that $y\ge z {\mathscr C}$, and the dual SDP becomes
\begin{equation}
-\min(r+z {\mathscr C})\quad\mbox{subject to}\;  R\ge0,
\label{dual simple}
\end{equation} 
where $R$ [the second constraint in Eq.~(\ref{supp DSDP})] can be written as $ R=H\!+\!(r\!-\!z)\openone
+z|\Psi^+\rangle\langle\Psi^+|$. 

Let us find the normalized eigenstate \mbox{$|\varphi\rangle\!:=\!\sum_{k=0}^n b_k|k\rangle$} 
of $ R$ associated to the minimum eigenvalue, i.e., the coefficients $\{b_k\}_{k=0}^n$, with $\sum_{k=0}^n b_k^2=1$, that minimize 
%
\begin{equation}
{\mathcal R}:=\langle\varphi| R|\varphi\rangle=\langle\varphi|H|\varphi\rangle+r-z+z|\langle\Psi^+|\varphi\rangle|^2
\label{min eig phi}
\end{equation} 
%
for given $r$, $z$. This will be done in two steps.
We will first constraint the state $|\varphi\rangle$ to satisfy the additional condition $\langle\Psi^+|\varphi\rangle:=\sum_{k=0}^nb_k=(n+2)^{1/2}\kappa=\kappa/\sqrt\epsilon$ for given $\kappa$. In a second step we will minimize over~$\kappa$. We can choose $\kappa\ge0$ without loss of generality. 

If $\kappa=0$, the minimization of ${\mathcal R}$ is straightforward, since the last term in Eq.~(\ref{min eig phi}) vanishes. One can verify that in this case the minimum is attained by~$|\varphi\rangle=|\psi_{\rm odd}\rangle$, where $|\psi_{\rm odd}\rangle$ is defined in Eq.~(\ref{odd state}) (MT). This state clearly has $\kappa=0$, since its components 
satisfy the relation \mbox{$b_k=-b_{n-k}$}. In this regime, we have 
\begin{equation}
\min_{|\varphi\rangle}{\mathcal R}\sim {4\pi^2\over n^2}+r-z.
\label{min 2}
\end{equation}

If $\kappa>0$, we immediately realize that the minimization of ${\mathcal R}$ was already carried out when we dealt with pure states in the MT, provided we drop the positivity condition, $b_k\ge0$, $k=0,1,\dots, n$ [i.e., provided $\sigma(t)\equiv0$ in Eq.~(\ref{S + lagrange}) of the MT], where now $b_{k-1}/\sqrt\epsilon$ and $\kappa$ play the role of $x(\epsilon k)$ and $l$, respectively. Reusing the results there, we have that the shape of $\{b_k\}_{k=0}^n$ for large $n$ is given by Eq.~(\ref{x omega}) (MT), and
\begin{equation}
{\mathcal R}\sim{2S(\omegap)\over n^2}+\left(n \kappa^2-1\right)z+r.
\end{equation}
The parameter $\omegap$, analogous to $\omega$, is such~that \mbox{$\kappa=f(\omega')$},
where the functions $f$ 
and $S$ are defined in Eq.~(\ref{f(w)}) and in the lines that follow in the MT. 
We now trade $\kappa$ for $\omega'$ and minimize over~$\omega'$.
The minimum of ${\mathcal R}$ is easily seen to be at a value of $\omega'$, $\pi\le\omegap$, satisfying 
\begin{equation}
n^3z\sim{\omegap^2\over1-(2/\omegap)\tan(\omegap/2)}.
\label{z omega}
\end{equation}
Note that $z=0$ implies $\omegap=\pi$, as it should be. Note also that because we dropped the positivity condition on $b_k$ [and, thus on $x(t)$], the parameter $\omegap$ is now unbounded from above. 
We will not attempt to invert Eq.~(\ref{z omega}), instead (and again) we parameterize everything 
in terms of~$\omegap$. Despite the cumbersome algebra involved, we obtain the remarkably simple result
\begin{equation}
\min_{|\varphi\rangle}{\mathcal R}\sim{\omegap^2\over n^2}+r-z ,
\label{min omega}
\end{equation}
%
where $z$ is given in terms of $\omega'$ by Eq.~(\ref{z omega}). 

\begin{figure}[ht]
\includegraphics[scale=.4]{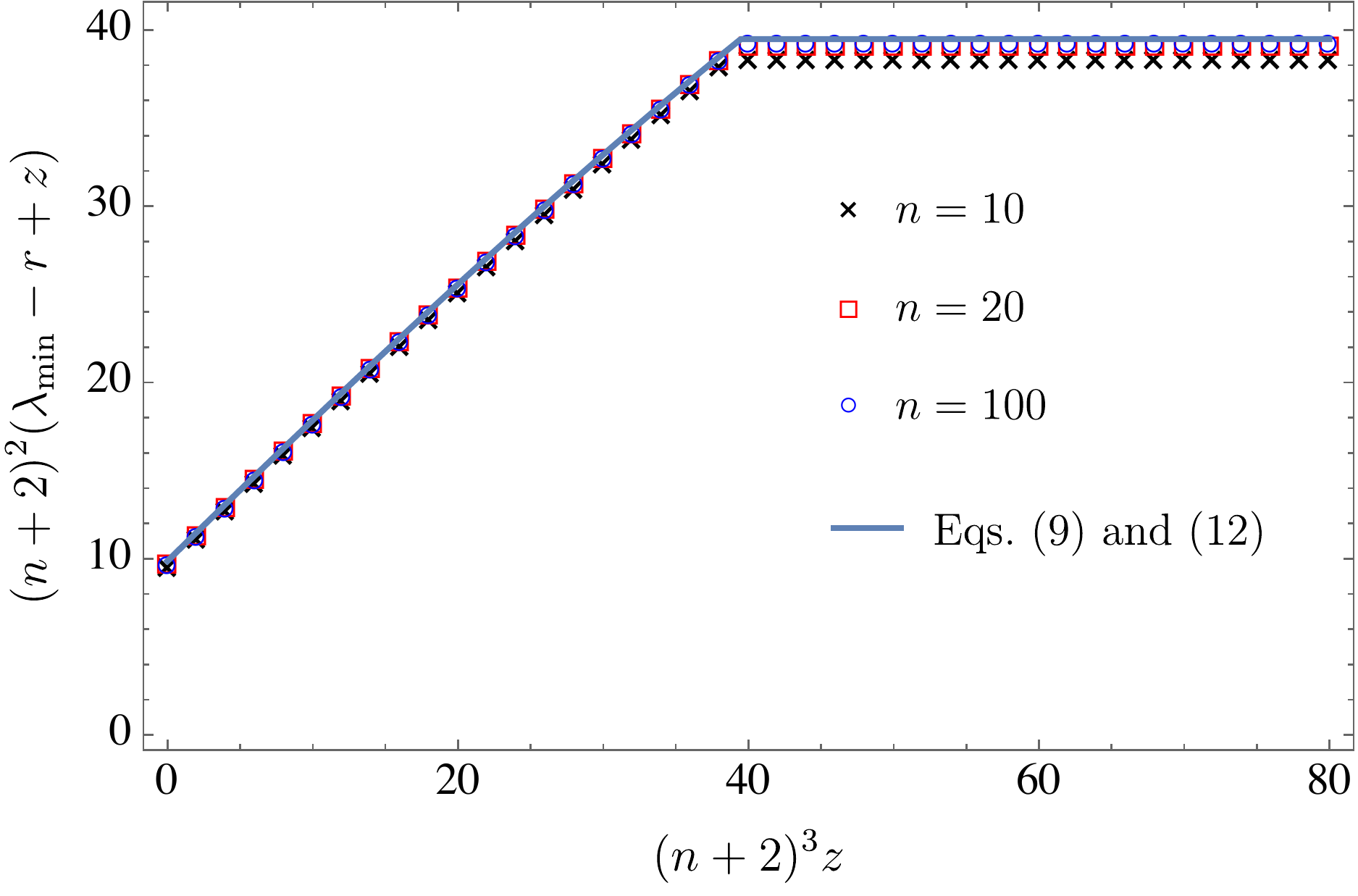}
\caption{Plot of $(n+2)^2(\lambda_{\rm min}-r+z)$ vs $(n+2)^3z$. The solid line is obtained by combining Eqs.~(\ref{min 2}) and~(\ref{min omega}). The black crosses, red squares and blue circles have been obtained numerically, for $n=10$, $20$ and $100$, respectively. The numerical points are seen to converge to the solid line as $n$ increases.
} 
\label{fig:5}
\end{figure}

A glance at Eqs.~(\ref{min 2}) and~(\ref{min omega}) shows that the former (latter) gives the true minimum eigenvalue of $R$, $\lambda_{\rm min}$, if it holds that $\omega'\ge2\pi$ (if $\pi\le\omega'\le 2\pi$). 
These expressions are compared to the  minimum eigenvalue of $R$, obtained numerically, in Fig.~\ref{fig:5}.
%

Let us first assume that~$\lambda_{\rm min}$ is given by Eq.~(\ref{min omega}).
Then, it follows from $R\ge 0$ that
\begin{equation}
r\sim{\omegap^2\over n^2}\left\{ {1/n\over 1-(2/\omegap)\tan(\omegap/2)}-1\right\}.
\end{equation}
Using this in Eq.~(\ref{dual simple}) we see that it just remains to minimize
\begin{align}
r+z{\mathscr C}\sim{\omegap^2\over n^2}\left[{l^2\over 1-(2/\omegap)\tan(\omegap/2)}-1\right],
\label{r+zC}\\[-1em] \nonumber
\end{align}
over $\pi\le\omegap\le2\pi$, where we recall that $l$ is defined in terms of~${\mathscr C}$ as $l=[({\mathscr C}+1)/(n+2)]^{1/2}$. Taking derivative with respect to $\omega'$, this straightforward unconstrained minimization leads to the condition $f(\omega')=l=f(\omega)$, where we have used Eq.~(\ref{f(w)}) (MT). 
In the given range of~$\omega'$, this last equation has the unique solution $\omega'=\omega$.
Substituting this value back in Eq.~(\ref{r+zC}) we obtain
\begin{equation}
-\min(r+z {\mathscr C})\sim{2S(\omega)\over n^2} .
\label{dual sol}
\end{equation}
This is the asymptotic solution for pure states obtained in the MT. Thus, we have shown that 
pure states are optimal in this regime of high coherence, or more precisely, they are optimal provided $c={\mathscr C}/{\mathscr C}^*\ge \pi^2[f(2\pi)]^2/8=\pi^2/12\approx0.822$.

Let us now assume that $\lambda_{\rm min}$ is given instead by Eq.~(\ref{min 2}). 
Then  $z\ge 4\pi^2/n^3$, where this lower limit is obtained by substituting \mbox{$\omega'=2\pi$} into the right hand side of Eq.~(\ref{z omega}). 
The positivity of $ R$ requires that $r\ge z-4\pi^2/n^2$, and
\begin{equation}
r+z {\mathscr C}\sim(1+{\mathscr C})z-{4\pi^2\over n^2},\quad z\ge {4\pi^2\over n^3}.
\end{equation}
With this, the minimum in Eq.~(\ref{dual simple}) gives the following lower bound to $\mathscr V$:
\begin{equation}
{4\pi^2\over n^2}\left(1-{{\mathscr C}+1\over n}\right)\le{\mathscr V},
\label{low bou -1}
\end{equation}
where the second term in the parenthesis cannot be discarded in the asymptotic limit because, as we recall,  ${\mathscr C}$  typically is $O(n)$. Actually, Eq.~(\ref{low bou -1}) can be more conveniently written for our purposes as
\begin{equation}
4\left(1-{8\over\pi^2}c\right)\le h.
\label{low bou}
\end{equation}
This lower bound is not attainable with pure probe states (except at the point $c=\pi^2/12$), as discussed in the MT and shown in Fig.~\ref{fig:4} (MT).

\subsection{\boldmath Suboptimality of pure states for $61.7\%\!<\!c\!<\!82.2\%$}

Let us consider the rank two state [Eq.~(\ref{the rho guess 2}) (MT)]
\begin{equation}
\rho={3l^2\over2}|\phi_{\rm even}\rangle\langle\phi_{\rm even}|\!+\!\left(\!1\!-\!{3l^2\over2}\right)|\phi_{\rm odd}\rangle\langle\phi_{\rm odd}|,
\label{rho guess 2}
\end{equation}
where the orthonormal states $|\phi_{\rm even}\rangle$ and $|\phi_{\rm odd}\rangle$ are defined as [Eqs.~(\ref{even state}) and~(\ref{odd state}) (MT)] 
\begin{align}
|\phi_{\rm even}\rangle&=\sqrt{8\over3(n+2)}\;\sum_{k=0}^n \sin^2\left(\pi{k+1\over n+2}\right)|k\rangle,\\[.2em]
|\phi_{\rm odd}\rangle&=\sqrt{2\over n+2}\;\sum_{k=0}^n \sin\left(2\pi{k+1\over n+2}\right)|k\rangle,
\end{align}
and we recall that $l=[({\mathscr C}+1)/(n+2)]^{1/2}$. Here, the subscript even (odd) refers to the symmetry of these states under the change $k \to n-k$. The function~$x(t)$ in Eq.~(\ref{x a}) is a continuous version of $|\phi_{\rm even}\rangle$ when $a=1$. We assume that $l\le f(2\pi)=\sqrt{2/3}$ (i.e., $c\le\pi^2/12\approx82.2\%$). The matrix $\rho$ is no longer positive semidefinite for larger values of $l$; we dealt with this situation in the previous section and proved that pure states are then optimal.
We take~$\rho$ as our optimal probe state guess for $l$ in the range~$[1/\sqrt2,\sqrt{2/3}]$.

We wish to show that ${\mathscr C}_R(\rho)={\mathscr C}$, however,  we will first use the next lemma to prove the weaker result~${\mathscr C}_R(\rho)\le{\mathscr C}$.

\begin{lem}\label{lemma 1} Let $\rho$ be the density matrix defined in Eq.~(\ref{rho guess 2}) and \mbox{$l\in[1/\sqrt 2,\sqrt{2/3}]$}, then
\begin{align}
&2l^2\sum_{k=0}^n \sin^2\left(\pi{k+1\over n+2}\right)|k\rangle\langle k|-\rho\ge0
\label{Pi 2}
\end{align}
for $n$ asymptotically large.
\end{lem}
%
From Lemma~\ref{lemma 1}, which will be proved in the next section, it follows that
\begin{equation}
\tau={2l^2\over{\mathscr C}}\sum_{k=0}^n \sin^2\left(\pi{k+1\over n+2}\right)|k\rangle\langle k|-{\rho\over{\mathscr C}}
\end{equation}
is a physical state such that $\rho+{\mathscr C}\tau$ is incoherent. Hence, according to the definition of RoC, Eq.~(\ref{def of RoC}) in the MT,
we have the bound ${\mathscr C}_R(\rho)\le{\mathscr C}$. 

Let us next compute the variance of $\rho$, namely, ${\mathscr V}(\rho)$, in the asymptotic limit. To this end, we need ${\mathscr V}(|\phi_{\rm even}\rangle)$ and~${\mathscr V}(|\phi_{\rm odd}\rangle)$. The first variance was already computed when we considered pure states in the MT, and we recall that ${\mathscr V}(|\phi_{\rm even}\rangle)\sim(4/3)\pi^2(n+2)^{-2}$. For large $n$ the components of $|\phi_{\rm odd}\rangle$ approach $x_{\rm odd}(t)=\sqrt2\sin(2\pi t)$ and  $S_{\rm odd}=(1/2)\int_0^1 \dot x_{\rm odd}^2(t)dt=2\pi^2$. We thus obtain the asymptotic expression ${\mathscr V}(|\phi_{\rm odd}\rangle)\sim4\pi^2(n+2)^{-2}$. By linearity, we have ${\mathscr V}(\rho)\sim 4\pi^2\!\left(1-l^2\right)(n + 2)^{-2}$. In terms of the ratios $c$ and $v$, the last expression can be written as
\begin{equation}
h\le{{\mathscr V}(\rho)\over {\mathscr V}^*}\sim 4\left(1-{8\over\pi^2}c\right).
\label{upp bou}
\end{equation}
Combining Eqs.~(\ref{upp bou}) and~(\ref{low bou}) we see that the latter bound is attainable for $\pi^2/16\le c \le\pi^2/12$, the state in Eq.~(\ref{rho guess 2}) is then optimal and its RoC must be~${\mathscr C}_R(\rho)={\mathscr C}$.

\subsection{Proof of Lemma~\ref{lemma 1}}

The proof of Lemma~\ref{lemma 1} is actually a calculation in the asymptotic limit of the lowest eigenvalue of the matrix 
in Eq.~(\ref{Pi 2}):
\begin{equation}
Q:=2l^2\sum_{k=0}^n \sin^2\left(\pi{k+1\over n+2}\right)|k\rangle\langle k|-\rho.
\end{equation}
%
Proceeding as we did in the MT and in previous sections of this supplemental material, we will find the state \mbox{$|\varphi\rangle=\sum_{k=0}^n b_k|k\rangle$} of unit norm that minimizes \mbox{${\mathcal Q}:=\langle\varphi|Q|\varphi\rangle$}. As $n$ goes to infinity, the components of~$|\varphi\rangle$ approach a function of a real variable $t$ that we denote as usual by~$x(t)$.  We will, thus, minimize
\begin{multline}
{\mathcal Q}=2l^2\int_0^1\sin^2(\pi t)x^2(t) dt-\\
{3l^2\over2}J^2_{\rm even}-\left(1-{3l^2\over2}\right)J^2_{\rm odd},
\label{obj func}
\end{multline}
over functions $x(t)$, normalized as
\begin{equation}
\int_0^1\!\! x^2(t)dt\!=\!1, 
\label{norm cond}
\end{equation}
where
\begin{align}
J_{\rm even}&=\int_0^1\sqrt{8\over3}\sin^2(\pi t) x(t) dt,\label{J even}\\[.2em]
J_{\rm odd}&=\int_0^1\sqrt{2}\sin(2\pi t) x(t) dt.\label{J odd}
\end{align}
We will carry out the minimization in two steps. First,~we will minimize over such function for given $J_{\rm even/odd}$, thus regarding Eqs.~(\ref{J even}) and~(\ref{J odd}) as constraints. Second, we will minimize over al possible values of $J_{\rm even/odd}$. We start by writing down an ``action" for $x(t)$:
\begin{multline}
S=2\int_0^1\left(l^2\sin^2(\pi t)+\mu^2\right)x^2(t) dt-2\mu^2\\
\kern-1em-2\sqrt3\,\alpha\left[\int_0^1\sqrt{8\over3}\sin^2(\pi t) x(t) dt-J_{\rm even}\right]\\
-2\beta\left[\int_0^1\sqrt{2}\sin(2\pi t) x(t) dt-J_{\rm odd}\right].
\label{new S}
\end{multline}
Here, we have introduced three \mbox{Lagrange multipliers}, $\mu^2$, $\alpha$, and $\beta$ (alongside with some convenient numerical factors $2$, $2\sqrt3$, and $2$), to enforce the constraints in \mbox{Eqs.~(\ref{norm cond})--(\ref{J odd})}. 
Solving the ``equation of motion" that follows from $S$ (just an algebraic equation because $S$ lacks a ``kinetic term'') we obtain
\begin{equation}
x(t)={2\alpha\sin^2(\pi t)+\beta\sin(2\pi t)\over\sqrt2\left[l^2\sin^2(\pi t)+\mu^2\right]}.
\label{x a b}
\end{equation}
It proves convenient to introduce the (polar) notation: $l=r\cos\theta$, $\mu=r\sin\theta$, where $0\le\theta\le\pi/2$.
The normalization condition, Eq.~(\ref{norm cond}), leads to
\begin{equation}
{\alpha^2(2+\sin\theta)+\beta^2\csc\theta\over r^4(1+\sin\theta)^2}=1.
\label{norm eq}
\end{equation}
Imposing Eqs.~(\ref{J even}) and~(\ref{J odd}) on $x(t)$, Eq.~(\ref{x a b}), 
one has:
\begin{align}
\alpha&={\sqrt3r^2 (1+\sin\theta)^2\over2(1+2\sin\theta)}J_{\rm even},\label{alph eq}\\[.2em]
\beta&={1\over2}r^2\left(1+\sin\theta\right)^2J_{\rm odd}.\label{beta eq}
\end{align}
By symmetry of $Q$ under ``parity", $k\to n-k$, its eigenstates must be either even or odd under the same transformation. This translates into $x(t)$ being either an even or an odd function in the interval $[0,1]$, and we just need to consider the two possibilities separately. Namely, $x=x_{\rm even}$ ($x=x_{\rm odd}$), for which $J_{\rm odd}=0$ ($J_{\rm even}=0$). 
Substituting Eqs.~(\ref{alph eq}) and~(\ref{beta eq}) in Eq.~(\ref{norm eq}) we obtain
\begin{align}
J_{\rm even}&={2(1+2\sin\theta)\over\sqrt3(1+\sin\theta)\sqrt{2+\sin\theta}}, & J_{\rm odd}&=0;\label{the J even}\\
J_{\rm odd}&={2\sqrt{\sin\theta}\over 1+\sin\theta }, & J_{\rm even}&=0\label{the J odd};
\end{align}
independently of $r$. This leads to:
\begin{align}
x_{\rm even}(t)&={\sqrt2(1+\sin\theta)\sin^2(\pi t)\over \sqrt{2+\sin\theta}\left[\sin^2\theta+\cos^2\theta\sin^2(\pi t)\right]},
\label{x even theta}\\
x_{\rm odd}(t)&={\sqrt{\sin\theta}\left(1+\sin\theta\right)\sin(2\pi t)\over \sqrt2\left[\sin^2\theta+\cos^2\theta\sin^2(\pi t)\right]}.
\label{x odd theta}
\end{align}
Now we are in the position to compute the first integral in Eq.~(\ref{obj func}), 
and after combining the resulting expression with Eqs.~(\ref{the J even}) and~(\ref{the J odd}), we obtain
\begin{align}
{\mathcal Q}_{\rm even}&=
{2l^2\sin^3\theta\over(2+\sin\theta)(1+\sin\theta)^2},\\
{\mathcal Q}_{\rm odd}&=2{l^2(4+\sin\theta)-2\over(1+\sin\theta)^2}\sin\theta.
\label{Pi odd theta}
\end{align}
It only remains to minimize over $0\le\theta\le\pi/2$. For the even state one can check that the minimum is attained at $\theta=0$, and from Eq.~(\ref{x even theta}) we have
$x_{\rm even}(t)=1$, with \mbox{$\min {\mathcal Q}_{\rm even}=0$}. For the odd state, the minimum of Eq.~(\ref{Pi odd theta}) is located at
\begin{equation}
\theta=\max\left\{0,\arcsin\left({1-2l^2\over 1-l^2}\right)\right\}.
\end{equation}
If $l\in[1/\sqrt2,\sqrt{2/3}]$ (the range we are interested in) one has $\theta=0$ and $x_{\rm odd}(t)=0$, for $t\in(0,1)$. In order to satisfy the normalization condition, Eq.~(\ref{norm cond}), $x^2_{\rm odd}(t)$ must be the distribution $x^2_{\rm odd}(t)= \delta(t)/2-\delta(1-t)/2$. Actually, one can easily check that for asymptotically large $n$, $b_0=-b_n\to 1/\sqrt2$, $b_k\to 0$, $k=1,2,\dots,n-1$ and $\min {\mathcal Q}_{\rm odd}\sim 2\pi^2l^2/n^2>0$. To summarize, in this range of $l$ the minimum eigenvalue of~$Q$ is $\lambda_{\rm min}=0$, and the statement of Lemma~\ref{lemma 1} holds true.

\begin{figure}[ht]
\includegraphics[scale=.21]{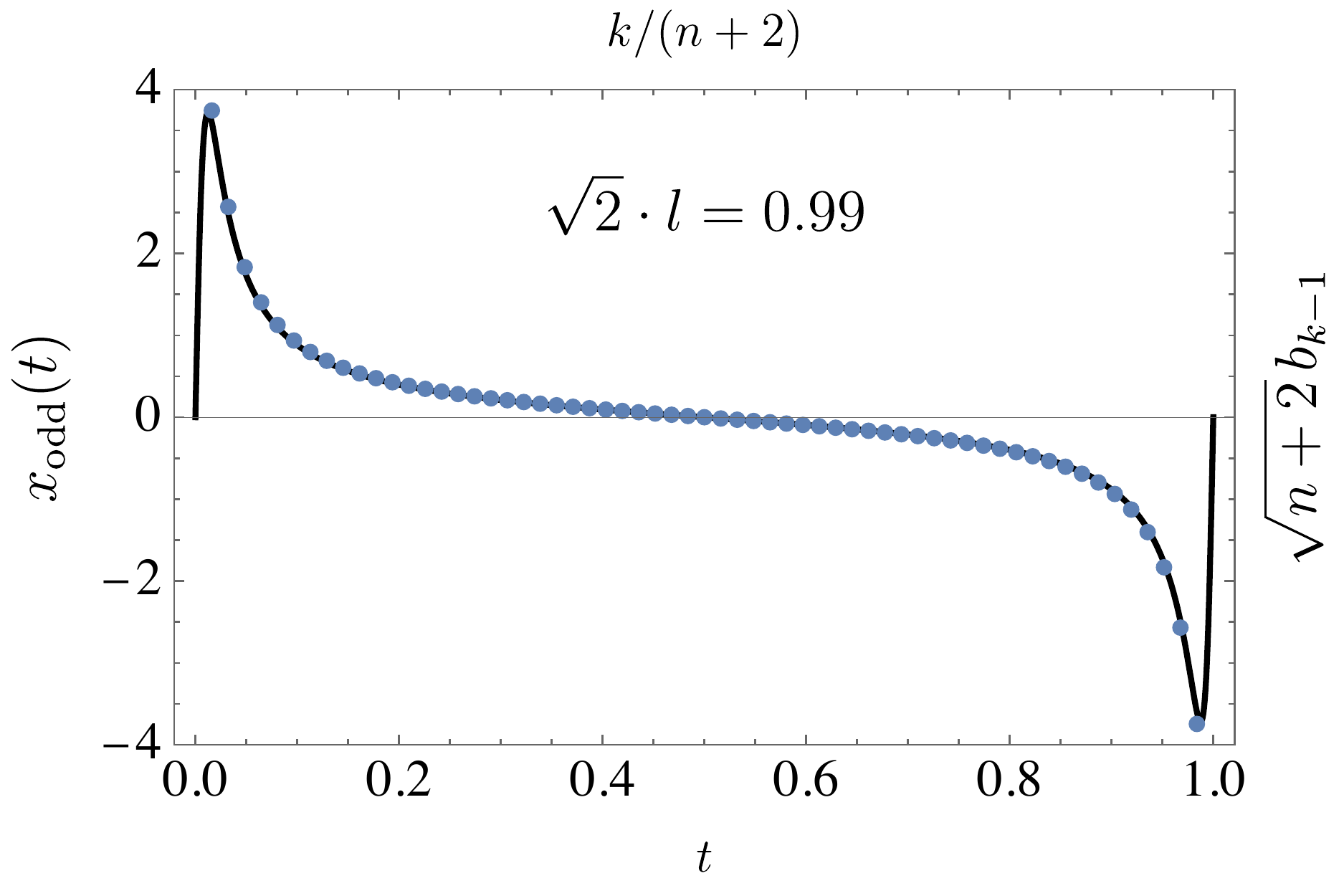}
\includegraphics[scale=.21]{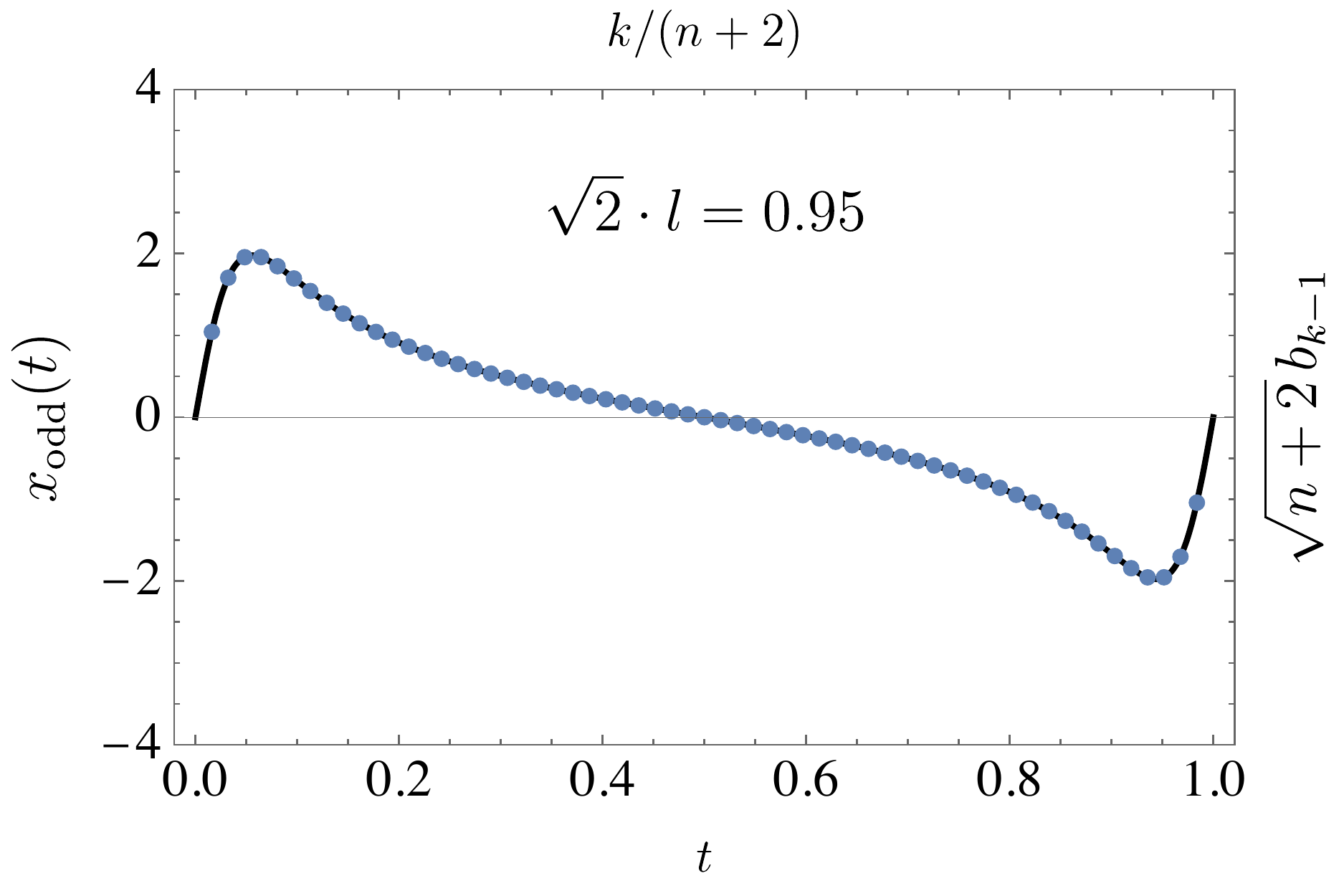}
\includegraphics[scale=.21]{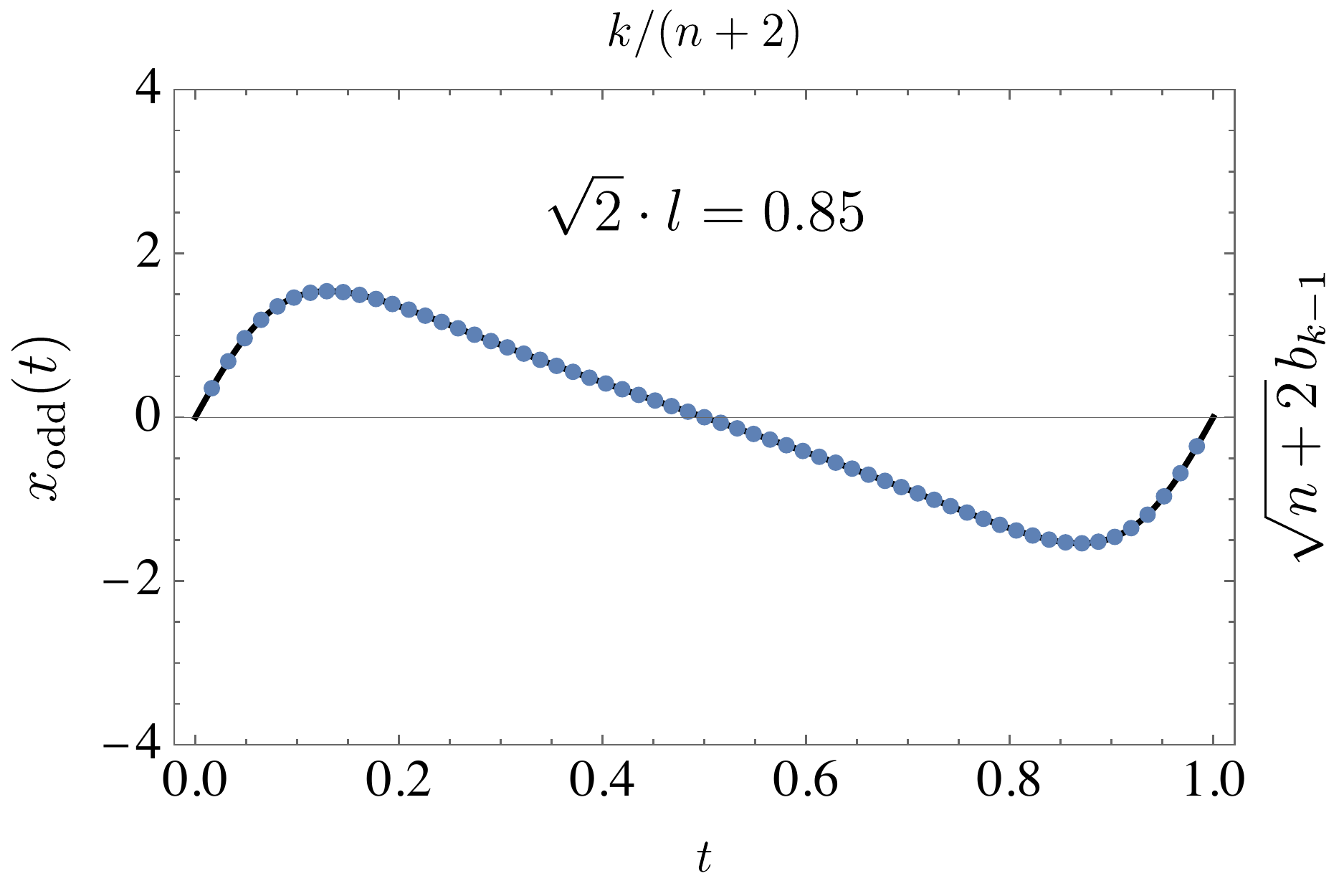}
\includegraphics[scale=.21]{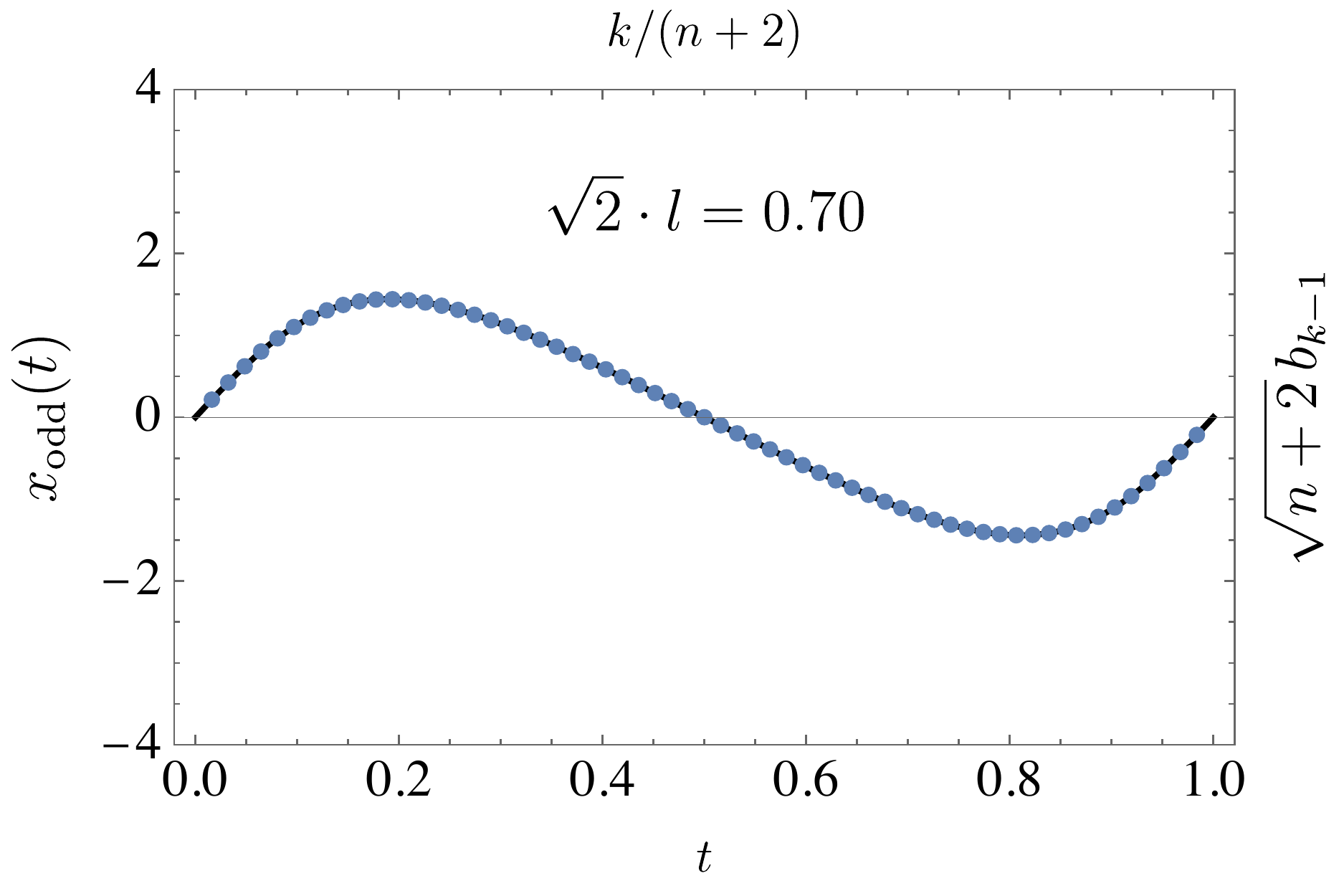}
\caption{In each plot, the solid line is the graph of the function $x_{\rm odd}(t)$, Eq.~(\ref{x_odd l}) (i.e., the eigenstate associated to the minimum eigenvalue of $Q$ in the asymptotic limit) for various values of $\sqrt2\cdot l$: $0.99$, $0.95$ (top row), and $0.85$, $0.7$ (bottom row). The dots are the points $(\epsilon k,b_{k-1}/\sqrt\epsilon)$, 
$k=1,2,\dots,n+1$, for $n=60$, where the components~$b_k$ of the eigenstate are computed by numerical methods.
} 
\label{fig:6}
\end{figure}

However, if $l\le1/\sqrt2$ the minimum of ${\mathcal Q}_{\rm odd}$ is attained at $\theta>0$. In this case one has
\begin{equation}
\lambda_{\rm min}\sim\min{\mathcal Q}_{\rm odd}=-{2(1-2l^2)^2\over2-3l^2}<0,
\end{equation}
and $Q$  ceases to be positive semidefinite, namely, Eq.~(\ref{Pi 2}) does no longer hold. In this regime the corresponding eigenvector can be rewritten as
\begin{align}
x_{\rm odd}(t)&={\sqrt{2(1-l^2)(1-2l^2)}(2-3l^2)\sin(2\pi t)\over2-6l^2+5l^4-l^2(2-3l^2)\cos(2\pi t)}.
\label{x_odd l}\\[-.7em]
& \nonumber
\end{align}
%

To check the validity of our approach, in Fig.~\ref{fig:6} we plot the function $x_{\rm odd}(t)$ along with the components, $b_k$, of the eigenvector with minimum eigenvalue that we have obtained by numerical methods for $n=60$. As one can see, the agreement is excellent even for this relatively small value of $n$.

\vfill

\end{document}